\documentclass[conference]{IEEEtran}
\usepackage{geometry}
\usepackage{graphicx}
\usepackage{cite}
\usepackage{tikz}
\usetikzlibrary{shapes,arrows}
\usetikzlibrary{calc}
\begin{document}

\title{\Large\bf High Area/Energy Efficiency RRAM CNN Accelerator with Kernel-Reordering
Weight Mapping Scheme Based on Pattern Pruning\\~\\}

\author{\normalsize
	\begin{tabular}[t]{c}
		\large Songming Yu, Yongpan Liu, Lu Zhang, Jingyu Wang, Jinshan Yue, \\
		\large Zhuqing Yuan, Xueqing Li, Huazhong Yang
		\\
		Department of Electronic Engineering, Tsinghua University, Beijing, China  \\
		e-mail: ypliu@tsinghua.edu.cn \\
	\end{tabular}}

\maketitle

\makeatletter
\def\ps@IEEEtitlepagestyle{%
  \def\@oddfoot{\mycopyrightnotice}%
  \def\@evenfoot{}%
}
\makeatother
\def\mycopyrightnotice{%
  \begin{minipage}{\textwidth}
    \footnotesize
    xxx-x-xxxx-xxxx-x/xx/\$31.00~\copyright~20xx IEEE \hfill\\~\\
  \end{minipage}
  \gdef\mycopyrightnotice{}
}

{\small\bf Abstract---Resistive Random Access Memory (RRAM) is an 
emerging device for 
processing-in-memory (PIM) architecture to accelerate convolutional
neural network (CNN).  However, due to the highly coupled crossbar structure
in the RRAM array, it is 
difficult to exploit the sparsity of the network in RRAM-based CNN 
accelerator. 
To optimize the weight mapping of sparse network in the RRAM array and
achieve high area and energy efficiency,
we propose a novel weight mapping scheme and 
corresponding RRAM-based CNN accelerator architecture 
based on pattern pruning and Operation Unit(OU) mechanism. 
Experimental results show that 
our work can achieve 4.16x-5.20x crossbar area efficiency, 1.98x-2.15x 
energy efficiency, and 1.15x-1.35x performance speedup in comparison with 
the traditional weight mapping method.
}

\section{Introduction}
In last decades, deep convolutional neural networks have been widely used
in many fields, such as face recognition, 
pose estimation, 
and image  classification\cite{ImageNet_DCNN}. Many CNN models can even achieve better accuracy than
humans\cite{lecun2015deep}. However, with the development of research, the sizes of modern 
CNN models are becoming larger, and every inference operation contains massive 
matrix-vector multiplications. 
The  computational complexity and storage overhead
of CNN models severely limit their application, such as in low-power embedded platform.

To overcome these drawbacks,
many dedicated hardware accelerator architectures have been proposed to accelerate
the neural network. 
Among them, 
RRAM-based CNN accelerator\cite{PRIME, ISAAC,PipeLayer} is one of the most 
promising architecture. It is a PIM architecture which uses RRAM array to perform 
arithmetic operations as well as data storage. 
This kind of accelerator can achieve high performance 
speedup and energy efficiency, because RRAM crossbar arrays can perform 
matrix-vector multiplication, which is  the most time and energy consuming operation 
in CNN, by means of high degree of parallelism and high efficiency.  

The RRAM has shown its great potential for accelerating the computation of the
neural networks, but the architectural exploration in 
the context of different application scenarios is still unknown. For example, 
the sparsity of the network has not been fully studied and optimized in RRAM-based
CNN accelerator. Previous works\cite{Channel-pruning,ADMM} 
have reported that there is significant weight redundancy
in neural networks. By properly designed fine-grained pruning algorithms, 
more than 85\% redundant weights 
in networks can be pruned with little accuracy loss\cite{structADMM}. However, this
kind of pruning algorithm makes the network structure irregular and aggravates the
high parallelism feature of the convolutional neural network, which severely affects 
the network mapping and computing efficiency in RRAM-based architecture. 
Due to the highly coupled RRAM crossbar structure, it is hard to optimize the 
weight mapping of irregular sparse networks to improve the area efficiency.
On the contrary, structured pruning, which
is hardware-friendly for RRAM-based architecture, causes more significant accuracy loss
than the fine-grained pruning with the same weight sparsity\cite{mao2017exploring}. 

Considering the drawbacks of previous pruning algorithms, recent 
researches\cite{ma2019pconv, pattern-pruning} 
mentioned about a new pruning algorithm---pattern pruning. 
Pattern pruning can be considered as an intermediate type between non-structured 
and structured pruning.
It can achieve both high accuracy, the feature of non-structured pruning, and high
regularity level, the feature of structured pruning. So, it is a very promising method
which contains the advantages of both non-structured and structured pruning.
We find it has the potential to be combined 
with RRAM-based CNN accelerator to optimize the weight mapping of spare network and
achieve high energy and area efficiency. Thus, we develop a novel weight mapping 
scheme for sparse network based on pattern pruning.
Besides, paper\cite{Sparse-ReRAM-Engine} mentioned that due to the hardware
limitation, the computation in the RRAM crossbar must be executed in smaller granularity,
called Operation Unit(OU), instead of activating an entire crossbar array in one cycle.
This mechanism also provides us more possibility to optimize the weight mapping of
sparse networks in RRAM array.

Our contributions mainly include: 
\begin{itemize}
	\item[1)] A novel area-efficiency weight mapping scheme based on pattern 
	pruning and OU mechanism to exploit the network weight sparsity in RRAM-based PIM architecture.
	\item[2)] An RRAM-based energy-efficiency sparse CNN accelerator architecture design 
	supporting our novel mapping scheme.
	\item[3)] Exploit both weight and feature map sparsity to achieve high energy
	efficiency in our architecture.
\end{itemize}

Experimental results show that in out pattern-pruned mapping
scheme, we can achieve 4.67x-5.20x crossbar area efficiency based on 
the pattern pruning results with almost no accuracy loss, which means
we save 76.0\%-80.8\% crossbar area than the naive mapping algorithm. And in 
our CNN accelerator design, we achieve a 1.98x-2.15x energy efficiency
and 1.15x-1.35x performance speedup.

\section{Background And Motivation}

\subsection{Weight Mapping in RRAM-based CNN Accelerator}
\begin{figure}[t]
	\centering
	\includegraphics[width = 0.4\textwidth]{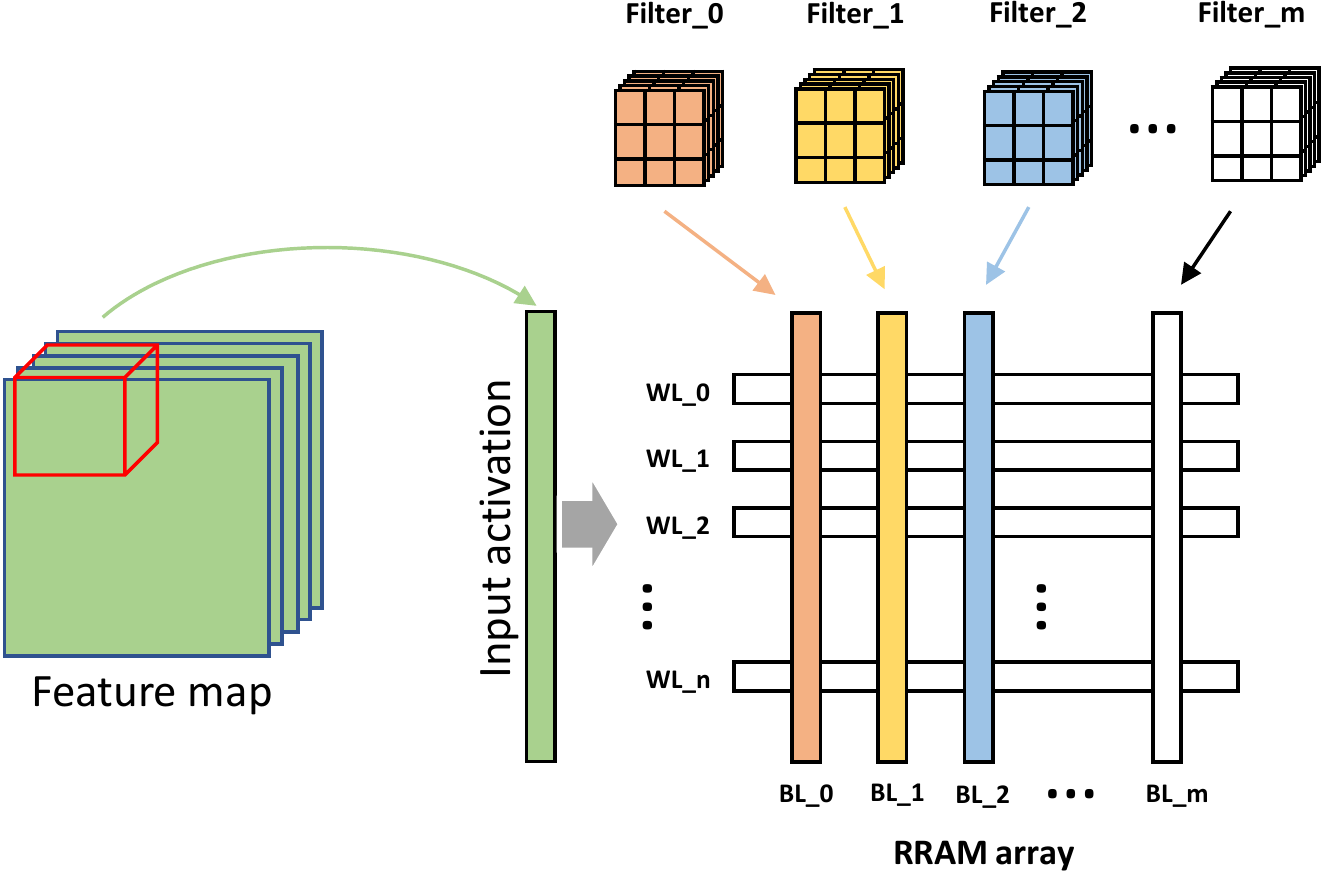}
	\caption{A naive weight mapping scheme(as a motivation example)}
	\label{naive_mapping}
\end{figure}

In RRAM-based CNN accelerator, RRAM crossbars are used to perform 
matrix-vector multiplication, which is the most energy and time consuming
operation in convolutional neural network. Figure \ref{naive_mapping} shows a naive 
weights mapping method for convolution layer weights. In previous
over-idealized RRAM-based CNN accelerator design\cite{PRIME,ISAAC,PipeLayer},
the whole crossbar is assumed to be activated in one clock cycle. And
all the weights in a filter are mapped to one column in the crossbar. 
When executing the matrix-vector multiplication, the input activations are 
converted to the voltage signals through DACs and 
fed into each row of the crossbar, then currents in each RRAM cell are accumulated
in the bitlines. Finally, the output current signals are collected 
at the end of each bitline and convert to digital signals by ADCs.  

However, this kind of mapping method and accelerator design makes it hard 
to exploit the sparsity of weights in RRAM-based CNN accelerator. 
In this method, all the weights in the same row of the RRAM array will be
multiplied with the same input, and the weights in the same column will accumulate to
the same output, so the position of the weights in the crossbar can't be easily changed.
If a weight is zero, it still needs to occupy an RRAM cell.
Only when all weights in a wordline or bitline are zeros can we remove
these zero weights and save those wordlines/bitlines. 

Furthermore, because of the conductance deviation per cell and the limitation
of the ADC resources, the activated wordline and bitline number in one cycle 
is limited. So only a small block of the RRAM crossbar, called operation unit\cite{Sparse-ReRAM-Engine},
can be executed per cycle.
For example, in a recent state-of-the-art RRAM-based CNN accelerator design\cite{ISSCC2018RRAM},
only nine wordlines and eight bitlines can be activated in a cycle. This
provides more possibility for us to exploit the weight sparsity of the convolutional 
neural network.

\subsection{Previous Works on Sparse CNN Accelerator}
Some studies have already realized the problem and proposed several solutions. 
\cite{ReCom} was aware of the problem of irregular sparsity, so
they turned to the regular sparsity and apply
regularization on filter and channel dimension to obtain structured sparse
network which can be directly mapped to the RRAM crossbar. 
But this paper does not mention the data about 
how much RRAM crossbar they save in their architecture. 
\cite{Sparse-ReRAM-Engine} makes use of the feature we mentioned above that 
only a small block, called Operation Unit(OU), can be executed per cycle. So 
they can exploit the weight sparsity in a small granularity. Compared to 
\cite{ReCom}, this paper achieves higher performance speech and energy saving.
But it does not mention the crossbar they saved either. 
\cite{10.1145/3287624.3287715} proposed a k-means clustering which shuffles the 
column in the weights matrix to gather the zero weights and use the 
crossbar-grained pruning algorithm to prune the all-zero crossbars. 
However, only 6\% to 22\% of crossbar resources can be saved in this algorithm,
which is still not efficient enough.

\begin{figure}[t]
	\centering
	\includegraphics[width = 0.4\textwidth]{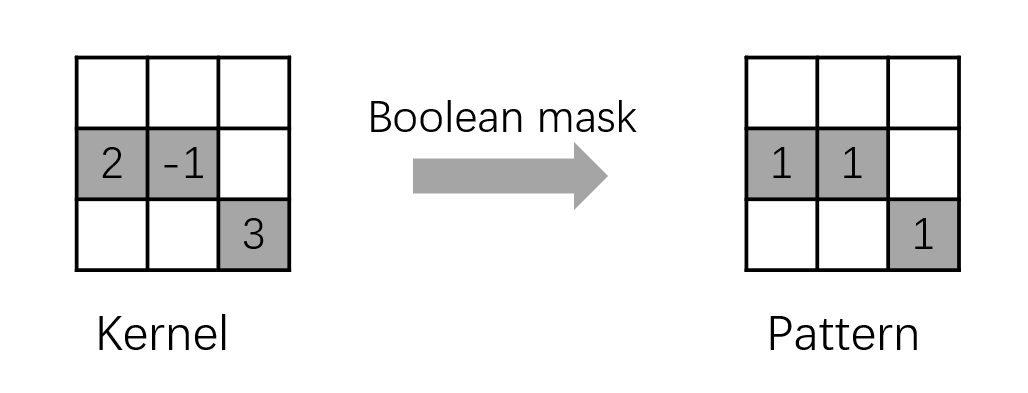}
	\caption{From kernel to pattern}
	\label{pattern}
\end{figure}

Recently, \cite{ma2019pconv,pattern-pruning} mentioned about a 
new kind of convolutional 
neural network pruning algorithm: pattern pruning. 
As figure \ref{pattern} shows, pattern means a shape
of kernels, and is defined as a boolean mask that indicates whether 
the weights are nonzero in each position. Taking the common $3\times3$
kernels as an example, in an irregular pruned network, weight in any position of
the kernel could be zero, so the theoretical maximum pattern numbers
is $ 2^{3\times3}$=512. By following proper pruning algorithm, we can limit the
total pattern numbers to a very low level, such as less than 8 patterns
in each network layer, with little accuracy and sparsity loss.
By pattern pruning, we can make the irregular sparse network
regular in kernel dimension.
Pattern pruning can achieve both accuracy and regularity at a higher level.
So we develop a novel weight mapping scheme based on pattern pruning to optimize
the mapping of the sparse network on the RRAM array and achieve high RRAM area
efficiency and energy efficiency.

\tikzstyle{startstop} = [rectangle,rounded corners,text centered, minimum width=3cm,minimum height=1cm,draw=black]
\tikzstyle{process} = [rectangle,text centered,text width =7cm,draw=black]
\tikzstyle{arrow} = [thick,->,>=stealth]

\begin{figure}[ht]
	\centering
	\begin{tikzpicture}[node distance=1.5cm]
		\node (start) [startstop] 
		{Pretrained network model};
		\node (process1) [process,below of=start] 
		{Pattern pruning: get a pattern-pruned network model};
		\node (process2) [process,below of=process1] 
		{Kernel reorder: gather the kernels with the same pattern};
		\node (process3) [process,below of=process2] 
		{Compress: remove zero elements and get compressed pattern block };
		\node (process4) [process,below of=process3] 
		{Pattern block placing: place the pattern blocks in the RRAM array properly};
		\node (stop) [startstop,below of=process4] 
		{Mapping finish};
		
		\draw [arrow] (start) -- (process1);
		\draw [arrow] (process1) -- (process2);
		\draw [arrow] (process2) -- (process3);
		\draw [arrow] (process3) -- (process4);
		\draw [arrow] (process4) -- (stop);

	\end{tikzpicture}
		
	\caption{The flowchart of our pattern-pruned weight mapping scheme}
	\label{pattern_pruned_mapping_workflow}
\end{figure}
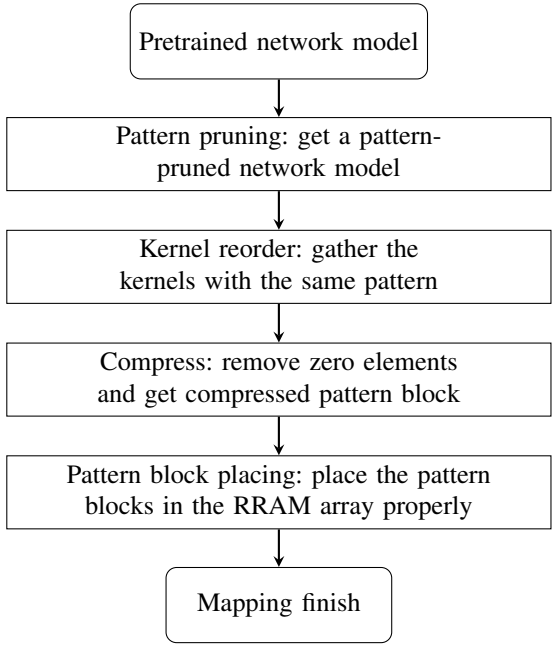
\section{Pattern-pruned Weight Mapping Scheme}

The flowchart in figure \ref{pattern_pruned_mapping_workflow} shows an overview
of our mapping scheme. We will explain each step in more detail in the following 
part of this section.

\subsection{Pattern Pruning}
Before we map the weights on the RRAM crossbars, we need to apply the
pattern pruning algorithm to get the pattern-pruned network. 

In pattern pruning algorithm, we start from an irregular pruned network.
First, we choose several most representative patterns in each layer, then
project other kernels to those chosen patterns by proper distance-based 
pattern projection method. Finally, we retrain the network to regain  
accuracy. 

In detail, we use the pruning algorithm in\cite{pattern-pruning}, an
alternating direction method of multiplier (ADMM)-based pattern
compression method. First, we calculate 
the probability density function (PDF) of all the patterns in the irregular 
pruned network. According to the PDF of the patterns, 
we choose several patterns
which have the largest probability as our candidate patterns. The number of 
candidate patterns we select is an adjustable parameter. The network is more
structured if the number of patterns in each layer is smaller, but the accuracy
will also decrease greater. So the number of patterns in each network layer is 
carefully chosen. After choosing the candidate patterns, we project other kernels
to the pattern in the candidate patterns which is closest to the original kernel.
The distance of patterns is the same as the vector distance, so we use the common
vector distance function, such as hamming distance, cosine distance, to measure
pattern distance. And the projection of kernels means element-wise multiplication
of the original kernels and the chosen pattern. After the projection, the network
is retrained to regain accuracy. The procedures above are repeated until the 
accuracy meets our expectation.

\begin{figure}[ht]
	\centering
	\includegraphics[width=0.48\textwidth]{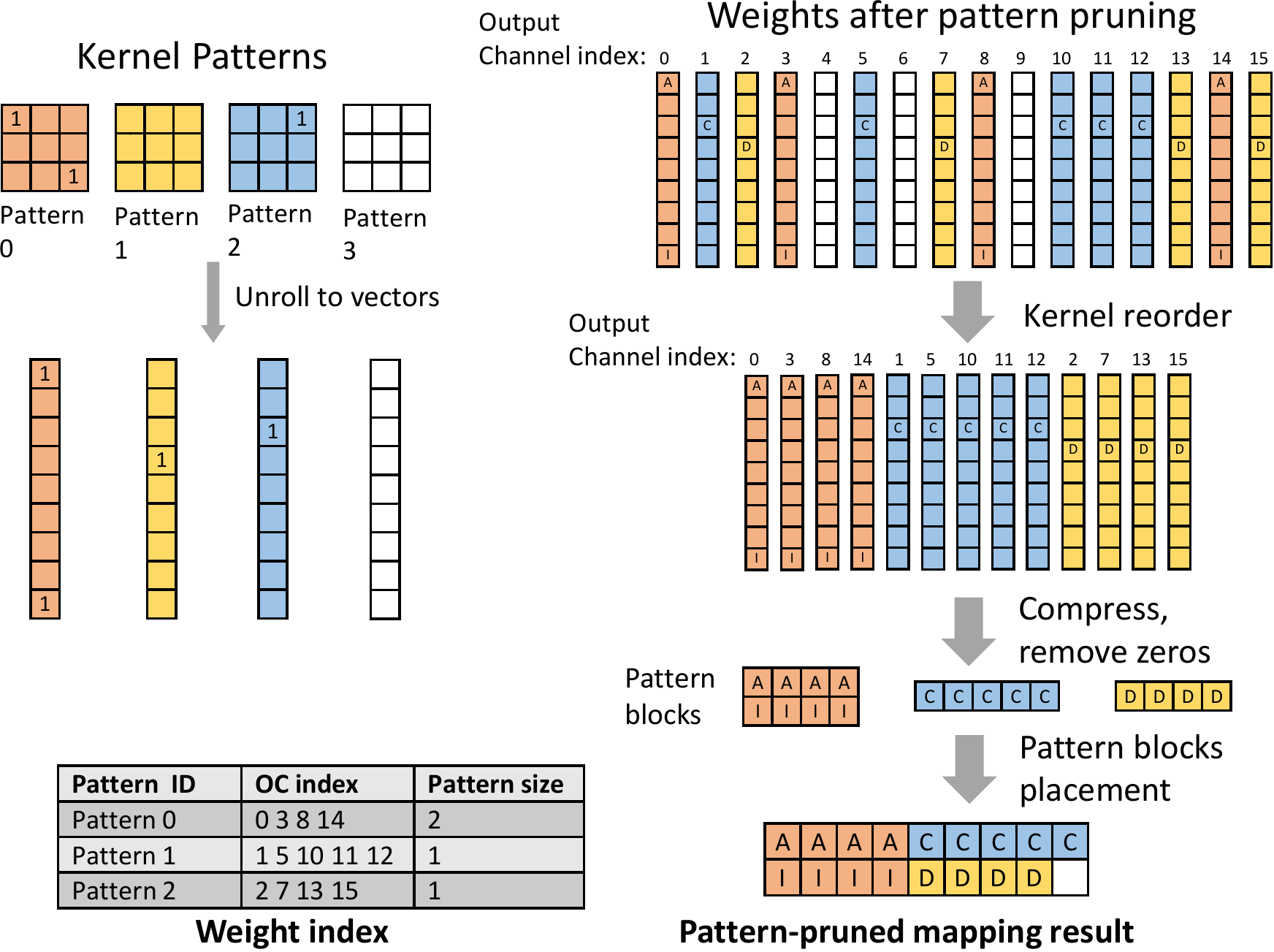}
	\caption{A case study of our pattern-pruned based mapping method. On
	the lower left is the weight index.}
	\label{pattern-pruned_mapping}
\end{figure}

\subsection{Weight Mapping Algorithm Workflow}
After we get the pattern-pruned network, we can perform our weight mapping 
algorithm.
Figure \ref{pattern-pruned_mapping} shows our mapping algorithm workflow.  
We take a small layer
with only one input channel and sixteen output channels for example. After
pattern pruning, all sixteen $3\times3$ kernels have only 4 patterns, including
an all-zero pattern. We unroll the patterns and kernels to one dimension vectors
and mark different patterns in different colors. Firstly, we reorder the
kernels according to the pattern types and gather the kernels with the same pattern.
Then, we can compress the kernels by removing all the zero elements.
After the compression, the adjacent kernels with the same pattern form a 
pattern block. Inside each pattern block, the matrix-vector multiplication can
be computed in parallel due to the fact that the weights in the same position of original
kernels are still in the same row in the crossbar. Finally, we place the 
pattern blocks on the crossbar by following a proper strategy. We will explain this strategy
in next paragraph.
In previous naive mapping
method, the weights matrix is directly mapped to the RRAM crossbar and
all those weights (16 kernels with the size of $3\times3$) will
take up a $9\times16$ crossbar array. But we optimize the pattern-pruned 
mapping scheme by storing all the weights in a $2\times9$ crossbar array. 

\begin{figure}[t]
	\centering
	\includegraphics[width=0.47\textwidth]{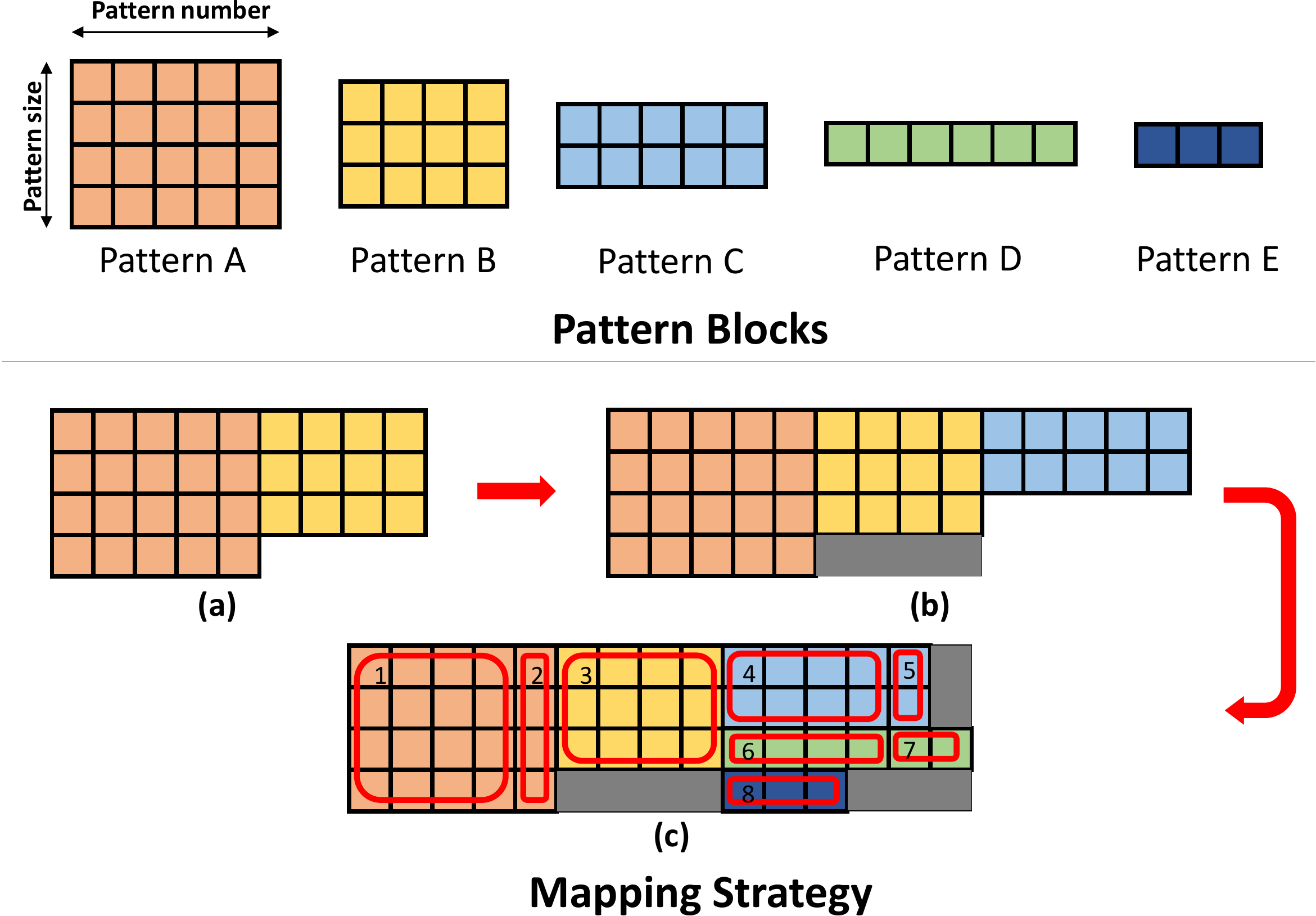}
	\caption{A more specific example to explain our mapping strategy.}
	\label{mapping-strategy}
\end{figure}

To explain how we place each pattern block on the crossbar (the last step in
figure \ref{pattern-pruned_mapping}), we use a more
specific example. As shown in figure \ref{mapping-strategy}, after getting 
each pattern block, we reorder all the blocks according to the pattern size
(the number of nonzero elements in that pattern).
First, we place the pattern block with the biggest pattern size, and put 
the next pattern block on the left, aligning it to the top of the former 
block. Then, if the number of rows behind the current block is enough for the 
next block, we can place it there and align it left.
Otherwise, we place the next block on
the left and is also aligned to the top of the former block.
There is only one row left behind the current block, as shown in figure \ref{mapping-strategy}(a),
which is not enough for the next pattern block 
with a pattern size of two, so the next pattern block is placed in new columns.
And one row marked in grey colors is wasted,
as shown in figure \ref{mapping-strategy}(b).
The next two blocks with only one row can be placed behind the
former block and are left-aligned, and a little more area marked in grey is 
wasted. And the red boxes in figure \ref{mapping-strategy}(c) show the OU organization for $4\times4$ OU size.

For a practical network layer with more than one input channel, we 
apply all those operations for every input channel and store all
the weights channel by channel.

Besides, we need to store the indexes of 
the kernels, since we have reordered the kernels. We store the indexes pattern
by pattern in the same order as mapping the pattern blocks to the crossbar, 
and for each pattern, we store the corresponding output channel
index of each kernel and the pattern shape (including pattern size).
We will explain how we can get the placement information of the weights 
from the  indexes in next section.
And the index overhead will be 
analyzed in section \ref{evaluation}.

\section{Architecture}
\begin{figure}[ht]
	\centering
	\includegraphics[width=0.48\textwidth]{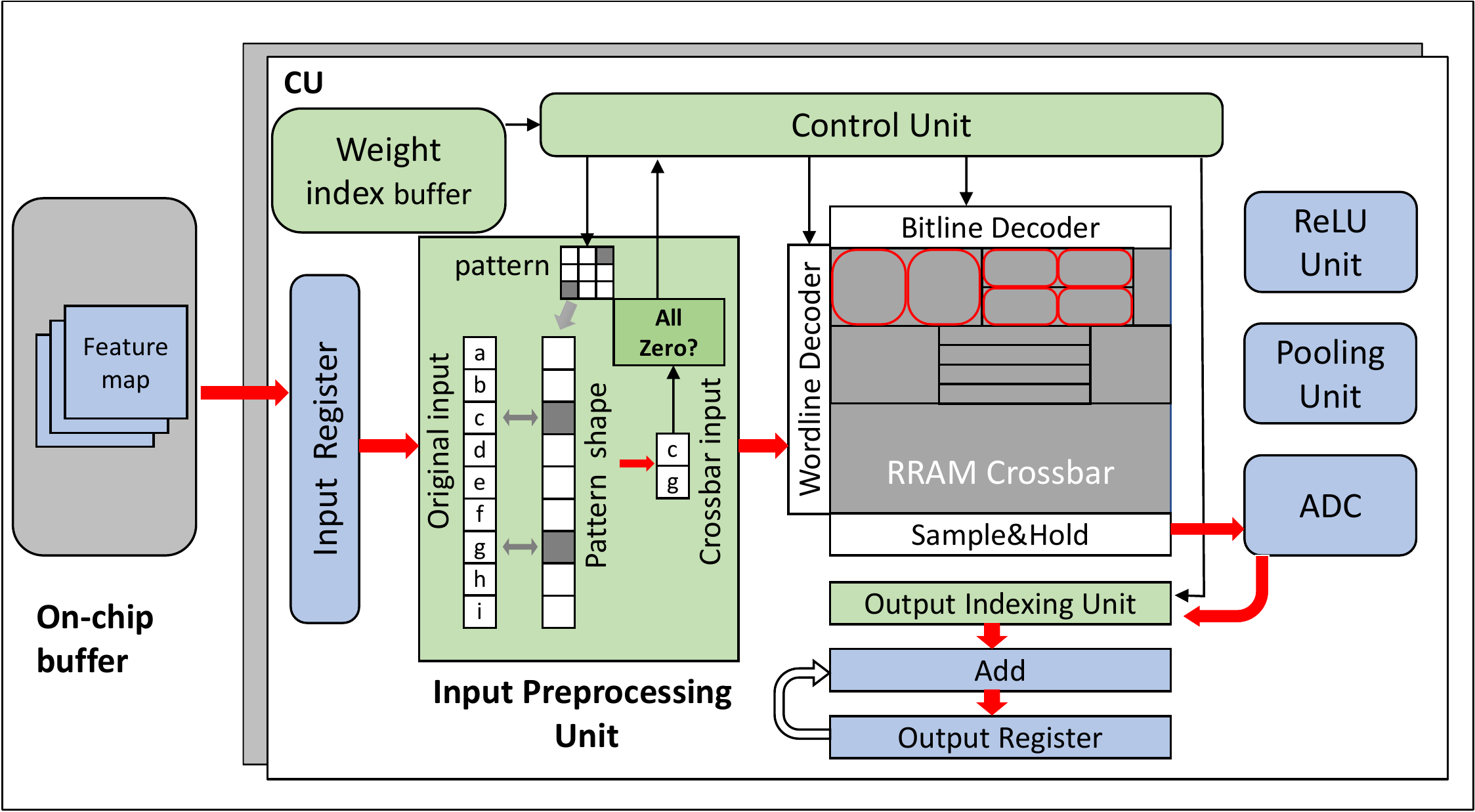}
	\caption{An overview of our RRAM-based CNN accelerator architecture.
	The red arrow shows the dataflow from input to output.}
	\label{architecture}
\end{figure}

In our mapping scheme, the weights in the RRAM crossbar are compressed and 
no longer in sequential order. So it cannot be deployed in traditional 
RRAM-based CNN accelerator without hardware design modification. 
Referencing former RRAM-based CNN accelerator architecture 
design\cite{PRIME,Sparse-ReRAM-Engine,10.1145/3316781.3317797}, we design 
a new architecture to support our mapping scheme. Figure \ref{architecture} shows 
our architecture design, and the red arrow shows the dataflow from
input to output. The weights are mapped to the RRAM crossbar by using our mapping
scheme, and the indexes are
stored in the weight index buffer. The main difference between the former 
designs\cite{10.1145/3287624.3287715,ISAAC}
and our architecture are explained in detail as follows. 

\subsection{Input Preprocessing Unit}
In the RRAM crossbar, we only store the nonzero weights. For 3*3 kernel, only 
nonzero elements are stored together, so when we send the inputs to crossbar,
we only send the input activations corresponding to the nonzero weights.
So we need to select the correct inputs according to the pattern of the current
weights. The Input Preprocessing unit is designed to implement this function.
It will get the pattern information from the control unit to send the 
inputs to the RRAM crossbar. Besides, we also notice that because of the ReLU 
activation function, there is also considerable sparsity in the input activations.
If we can utilize it, we can make further improvement in energy efficiency. So
we add an all-zero detection module in the Input Preprocessing unit. If the inputs are all
zeros, a signal is sent to the control unit and all the operations will not
be done to avoid useless computation and save energy.

\subsection{Output Indexing Unit}
In our mapping method, the weight kernels in the RRAM crossbar are no longer 
sequentially stored in 
each bitline of the crossbar, as we have explained in previous section.
And the outputs collected in each bitline are also out of sequence. So before
we store the outputs into the output register, we need to reorder those outputs.
In the output indexing unit, the outputs collected in every cycle are reordered 
and stored into the right address according to the indexes stored in the weight 
index buffer.

\subsection{Operation Unit Organization }
As we have mentioned before, in every cycle we can only activate a small block
of the RRAM crossbar (called operation unit) to perform the computation, instead 
of activating the whole crossbar. 
In the pattern-pruned weight mapping scheme, every operation unit
must be limited inside a pattern block, because
for different patterns, the weights stored in the same wordline correspond to
different inputs, and cannot be computed in parallel. 
In figure \ref{mapping-strategy}(c),
the red boxes show an example of OU Organization for an OU size of $4\times4$.

Another problem is how we can know the placement of the weights in the crossbar 
from the indexes.
Actually, the procedures are similar to the mapping strategy shown in figure \ref{mapping-strategy}. 
First, we get the index
of the pattern with the biggest pattern size, the width of this pattern block is
the number of input channels of this pattern, and the height of this pattern block
is the pattern size. We know the next pattern block is next to the current 
block, since this is how we place the blocks. We can also know the width and the
height of this pattern from the input channel indexes and the pattern size.
Then we get the pattern size of 
the next pattern block. If there are enough rows behind the current block for next
block, then we know it is placed there, otherwise we know that it will be placed
in new columns. 
Repeat those steps until we get all the weights' placement.

\section{Evaluation}
\label{evaluation}
\subsection{Evaluation Setup}
For hardware energy model settings, according to ISAAC\cite{ISAAC}, RRAM related
components (crossbars, ADCs, and DACs) consume more than 80\% energy of the total
chip, so we focus on those components when evaluating the energy efficiency. 
Table \ref{Hardware_Parameters} shows our hardware configuration.  For 
ADCs and DACs energy, we use the data from\cite{AERIS}. And the RRAM crossbar
array energy model is based on\cite{hu2016dot}. The Operation Unit size is set to
$9\times8$, the same as\cite{ISSCC2018RRAM}, 
which means that we can activate up to 9 wordlines and 8 bitlines per cycle.

\begin{table}[ht]
	\caption{Hardware Parameters}
	\begin{minipage}{8cm}
	\centering
		\def\arraystretch{1.5}\tabcolsep 2pt
		\def\thefootnote{a}\footnotesize
		\begin{tabular}{|c@{~~~}|c@{~~~}|c@{~~~}|c@{~~~}|}
			\hline
			\textbf{Components} & \textbf{Parameters} &\textbf{Spec}&\textbf{energy}\\
			\hline
			ADC        &Precision      & 8 bits   &1.67 pJ/op\\
					   &Frequency      & 1.2 GSps  &\\
			\hline
			DAC        &Precision      & 4 bits   &0.0182 pJ/op\\
			           &Frequency      & 18 MSps  &\\
			\hline
			RRAM Array &OU size        & $9\times8$ &4.8 pJ/OU/op\\
			           & bits per cell & 4          &  \\
			           & size          & $512\times512$    & \\
			\hline
		\end{tabular}
	\end{minipage}
	\label{Hardware_Parameters}
\end{table}

We use a modified VGG16 network as our benchmark. The convolution layers in our
network are the same as\cite{simonyan2014very}, but our network only contains one
full-connected layer. By modification, we greatly reduced the parameters in FC
layers, so we can focus on the results of the convolution layers. 
The datasets we use include CIFAR-10, CIFAR-100\cite{krizhevsky2009learning} and 
ImageNet\cite{deng2009imagenet}.

For comparison, we use the naive mapping method in figure \ref{naive_mapping}
as our baseline.
We build a simulator in Python to implement our mapping algorithm and simulate
the weight mapping workflow in the crossbars and computation workflow to get 
the crossbar area and energy efficiency and speedup.

\subsection{Pattern pruning result}

\begin{table*}[ht]
	\caption{Pattern Pruning Results}
	\centering
		\def\arraystretch{1.5}\tabcolsep 2pt
		\def\thefootnote{a}\footnotesize
		\begin{tabular}{c@{~~~~~~~}c@{~~~~~~~}c@{~~~~~~~}c@{~~~~~~~}c@{~~~~~~~}c@{~~~~~~~}}
			\hline
			\textbf{Dataset} & \textbf{Sparsity} &\textbf{Pattern Numbers in Each Conv layer}&\textbf{Total}&\textbf{top-1}&\textbf{top-5} \\
			\hline
			CIFAR-10        &86.03\%(+4.08\%)  & [2, 2, 2, 6, 8, 8, 8, 6, 5, 4, 6, 6, 8] &71   &92.63\%(-0.09\%)  &/  \\
			\hline
			CIFAR-100       &85.23\%(+3.28\%)  & [2, 2, 2, 2, 2, 8, 8, 8, 5, 6, 7, 6, 8] &66   &72.73\%(+0.01\%) &92.23\%(+0.79\%)  \\
			\hline
			ImageNet     &82.48\%(-0.90\%)    & [2, 2, 2, 2, 2, 9, 12, 12, 9, 10, 6, 4, 4] & 76 &71.15\%(-0.75\%)   &89.98\%(-0.51\%)  \\
			\hline
		\end{tabular}
	\label{Pattern_results}
\end{table*}

The baseline VGG16 networks are trained on CIFAR-10, CIFAR-100 and 
ImageNet, and are irregularly pruned\cite{ADMM}. The network trained on 
CIFAR-10 is of 81.95\% sparsity, 419 mean pattern numbers per layer 
and 91.72\% top-1 accuracy. The network trained on CIFAR-100 is of 
81.95\% sparsity, 450 mean pattern
numbers per layer and 72.72\%/91.44\% top-1/top-5 accuracy. The networks trained
on ImageNet is of 83.38\% sparsity, 461 mean pattern numbers per layer and 
71.90\%/90.49\% top-1/top-5 accuracy.

By pattern pruning, we can achieve 2-12 patterns per convolution layer, 
and more than 80\% sparsity in convolution layers. Table \ref{Pattern_results}
shows the pattern pruning results. After pattern pruning, the sparsity of the 
networks trained on CIFAR-10 and CIFAR-100 is even higher than the baseline 
networks, with little or no accuracy loss.
And the pattern numbers in each layer are no more than 12, while on CIFAR-10 and 
CIFAR-100 the numbers are no more than 8, makes the networks more structured.

\subsection{Evaluation Results}

\begin{figure}[t]
	\centering
	\includegraphics[width = 0.45\textwidth]{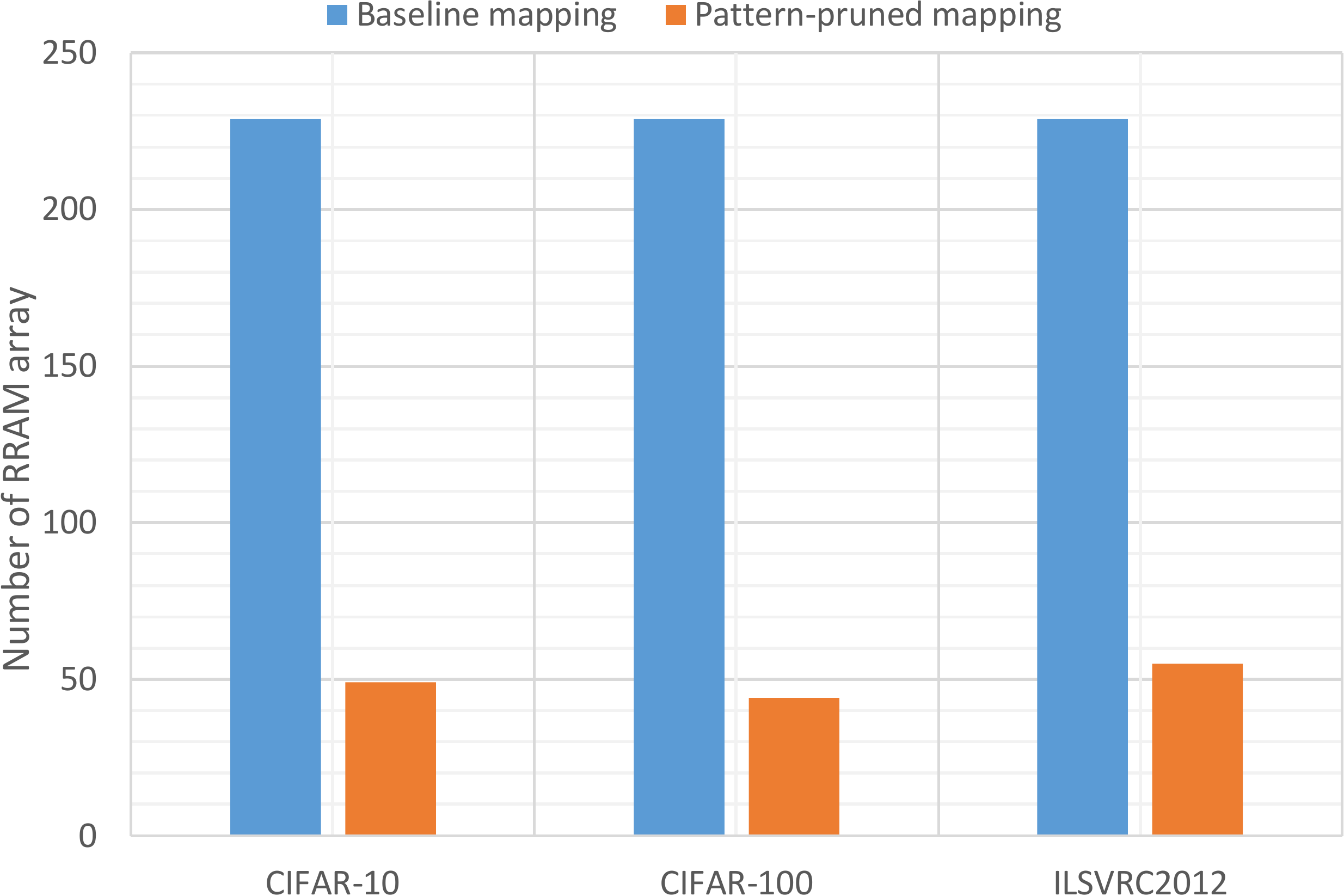}
	\caption{The results of RRAM crossbar array area efficiency on 
	different datasets. The y-axis is the crossbar array numbers used to
	store the network weights}
	\label{area_efficiency}
\end{figure}

\textbf{Crossbar Area Efficiency:} Figure \ref{area_efficiency} shows the results of
RRAM crossbar area efficiency. In our pattern pruned
mapping algorithm, we achieve an area efficiency improvement of $4.67\times$/
$5.20\times/4.16\times$ for networks trained on CIFAR-10, CIFAR-100, and ImageNet, respectively, 
which means we save 78.5\%/80.8\%/76.0\% RRAM crossbar array comparing to the baseline
mapping method. This is very close to the theoretical best results (86.03\%/85.23\%/82.48\%, 
the sparsity of the networks). This means our mapping algorithm has utilized
most of the sparsity of the networks. The result of the network trained in ImageNet
is relatively worse, and the main reason is that the pattern pruning result on this
dataset is not so good, just as table \ref{Pattern_results} shows. The sparsity of the network  
trained on ImageNet is lower than the other two, and the pattern numbers are relatively
higher, which means the network structure is more irregular. Lower sparsity and more
irregular structure make the mapping efficiency lower than others.

\begin{figure}[t]
	\centering
	\includegraphics[width = 0.45\textwidth]{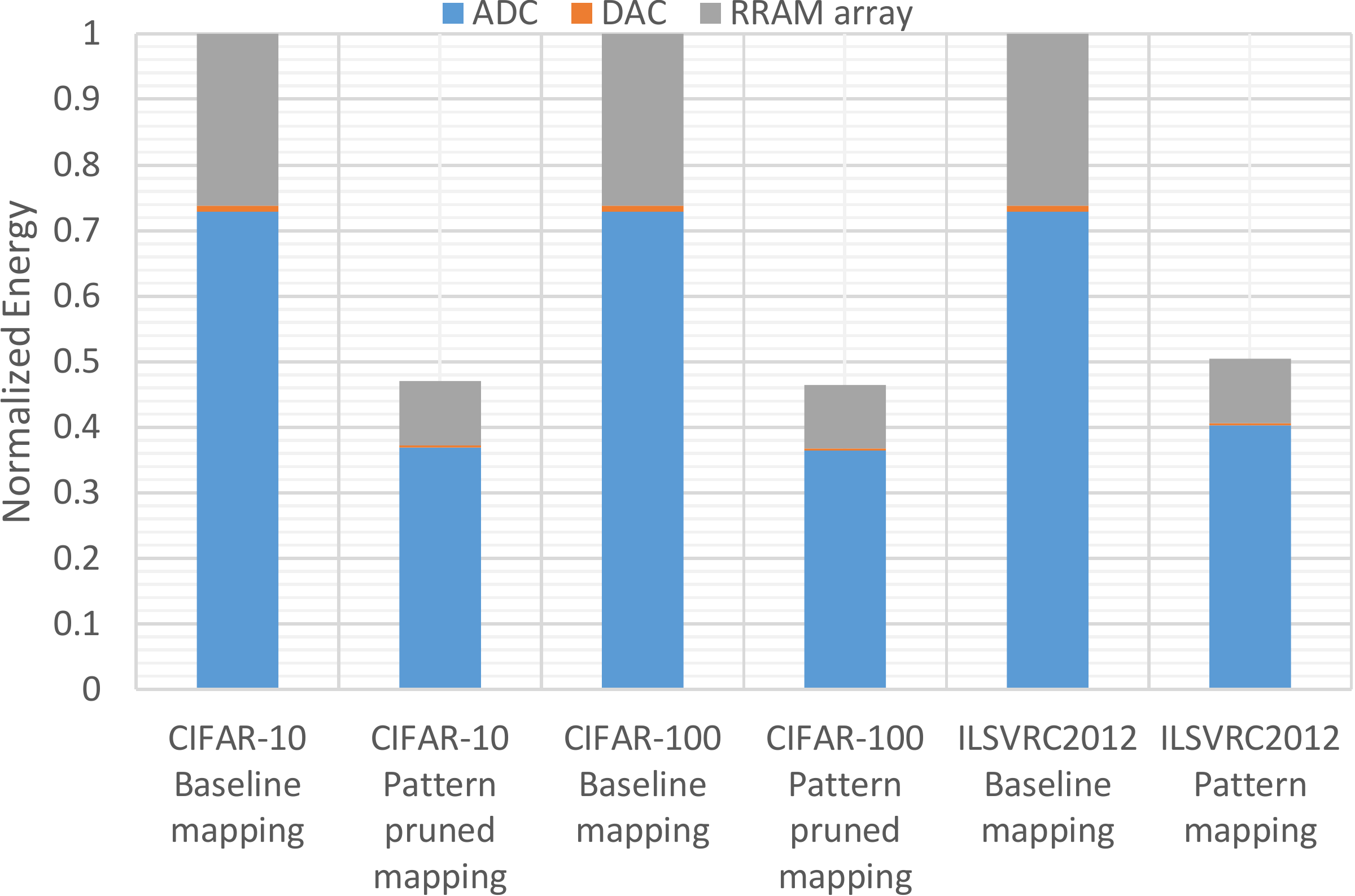}
	\caption{The results of normalized energy  on 
	different datasets. The energy data are all normalized to the baseline results.}
	\label{energy_efficiency}
\end{figure}

\textbf{Energy Efficiency:} By pattern-pruned mapping algorithm, much fewer
RRAM crossbar arrays are used, and in every cycle, less bitlines and wordlines,
as well as the ADCs and DACs, are activated because of the pattern pruned 
compression, so the energy efficiency is also higher than the baseline 
mapping algorithm. Besides, all-zero detection module in the Input Preprocessing unit also
makes important contribution to the improvement of energy efficiency.
Figure \ref{energy_efficiency} shows the energy efficiency results. We can see 
that the ADC energy is the main bottleneck. We achieve
$2.13\times/2.15\times/1.98\times$ energy efficiency on CIFAR-10, CIFAR-100, 
and ImageNet, respectively.

\textbf{Performance Speedup:} The speedup is achieved mainly by the deleted
all-zero patterns which are neither stored in crossbars nor computed. Though the 
speedup ratio is relatively small, only $1.35\times/1.15\times/1.17\times$ on CIFAR-10, 
CIFAR-100 and ImageNet respectively, it is acceptable
for us since we have achieved very high crossbar area efficiency and energy 
efficiency.

\subsection{Index overhead analysis}
As we have mentioned before, we need to store the kernels' output channel
indexes because we reorder the kernels inside every input channel. We also 
store the pattern shapes for each layer, but this overhead can be ignored 
compared to the
kernel indexes. For every kernel stored in the crossbars, we need an output
channel index with no more than 9 bits (for 512 output channels). 

In our evaluation, the total index overhead of the networks trained on CIFAR-10/CIFAR-100/ImageNet 
is 729.5KB/1013.5KB/990.6KB, respectively. 
The main factor that influences the size of the index overhead is the all-zero
pattern ratio in each network. And in our results, the all-zero pattern ratio 
in each network is 40.9\%/27.4\%/28.5\%. In our mapping method, 
all-zero patterns will not be stored in the crossbars, so their indexes will
also be saved. Compared to the total network size, the overhead of the indexes
is totally acceptable. For example, the size of the network trained on CIFAR-10
is 28.1MB before pruning, 6.0MB after pattern-pruned mapping(16 bits per weight),
so the index overhead is only 12.2\% of the network model size.

\section{Conclusions}
RRAM is an emerging device for PIM architecture and has shown its 
great potential for accelerating the
neural networks. To exploit the sparsity of the CNN
in the RRAM-based CNN accelerator, in this paper, we proposed a 
novel weight mapping scheme based on pattern pruning
and a corresponding CNN accelerator architecture design to support the
pattern-pruned weight mapping scheme.
The results of our experiment show that in out pattern-pruned mapping
scheme, we can achieve 4.16x-5.20x crossbar area efficiency based on 
the pattern pruning results with almost no accuracy loss, which means
we save 76.0\%-80.8\% crossbar area than the naive mapping algorithm. And in 
our CNN accelerator design, we achieve a 1.98x-2.15x energy efficiency
and 1.15x-1.35x performance speedup.

\bibliographystyle{ieeetr} 
\bibliography{ref}

\end{document}